\documentclass{ifacconf}

\usepackage{graphicx}      
\usepackage{natbib}        

\usepackage[dvipsnames,table]{xcolor}

\usepackage{amsmath,amssymb,amsfonts}
\newcommand*{\maxOp}{\operatornamewithlimits{max}\limits}

\usepackage[algo2e,ruled]{algorithm2e}

\usepackage{booktabs}

\usepackage{siunitx}

\begin{document}
\begin{frontmatter}

\title{Data-Driven Process Optimization of\\ Fused Filament Fabrication based on\\ \textit{In Situ} Measurements\thanksref{footnoteinfo}} 

\thanks[footnoteinfo]{Work supported by the Swiss Innovation Agency (Innosuisse, grant \textnumero 102.617) and by the Swiss National Science Foundation under NCCR Automation (grant \textnumero 180545).
\textit{E-mails:} \texttt{\{xaguidetti,ebalta,ralisa,lygeros\}@control.ee.ethz.ch}, \texttt{mkuehne@student.ethz.ch}, \texttt{yannick.nagel@nematx.ch} }

\author[ifa,inspire]{Xavier Guidetti},
\author[itet]{Marino Kühne}, 
\author[nematx]{Yannick Nagel},
\author[ifa]{Efe C. Balta},
\author[ifa,inspire]{Alisa Rupenyan},
\author[ifa]{John Lygeros}

\address[ifa]{Automatic Control Laboratory, ETH Zürich, Zürich, Switzerland }
\address[inspire]{Inspire AG, Zürich, Switzerland}
\address[itet]{D-MAVT, ETH Zürich, Zürich, Switzerland }
\address[nematx]{NematX AG, Zürich, Switzerland }

\begin{abstract}                
The tuning of fused filament fabrication parameters is notoriously challenging. We propose an autonomous data-driven method to select parameters based on \textit{in situ} measurements. We use a laser sensor to evaluate the surface roughness of a printed part. We then correlate the roughness to the mechanical properties of the part, and show how print quality affects mechanical performance. Finally, we use Bayesian optimization to search for optimal print parameters. We demonstrate our method by printing liquid crystal polymer samples, and successfully find parameters that produce high-performance prints and maximize the manufacturing process efficiency.
\end{abstract}



\end{frontmatter}

\section{Introduction}

In fused filament fabrication (FFF), the selection of optimal process parameters is a complex task that has been tackled with different methods~\citep{dey2019systematic}. The properties of a manufactured part are strongly dependent on a large number of inputs. This is particularly true for high-performance feedstock materials such as liquid crystal polymers (LCPs). Under- or over-extrusion, which are caused by poor parameter selection, have been shown to strongly influence the mechanical properties of polymers~\citep{siqueira2017}. As the evaluation of a sample surface can reveal print quality, we propose to use \textit{in situ} measurements of surface roughness to optimize the FFF process parameters. We do so by utilizing Bayesian optimization (BO), a sample-efficient approach to the optimization of problems that can be evaluated only point-wise. In Sec. \ref{sec:background} we introduce the literature and concepts upon which this work is based. Sec. \ref{sec:problem} formalizes the problem we tackle. Finally, Sec. \ref{sec:methods} and Sec. \ref{sec:res} present our approach to FFF parameters optimization based on \textit{in situ} data and the corresponding results.

\section{Background}\label{sec:background}

\subsection{FFF Parameters Configuration}

FFF has a large number of tunable process parameters that affect manufactured part properties such as dimensional accuracy, part strength, and surface roughness, among others~\citep{turner2015review}. A number of process optimization methods have been proposed in the literature, see~\citep{dey2019systematic} for a recent survey. Most of the existing methods rely on a design of experiments study based on e.g. Taguchi analysis~\citep{sood2009improving,wankhede2020experimental}, Q-optimal based models of experimental data~\citep{mohamed2016mathematical}, response surface methods, fuzzy inference systems, artificial neural networks, genetic algorithms~\citep{peng2014process}, and more.
However, these methods tend to be too specific and resource-intensive (e.g.\ using destructive testing) to characterize performance under changing process parameters. Additionally, the models often do not make use of \textit{in situ} measurements for the efficient evaluation of printed part characteristics.
In practice, there is a need for autonomous optimization methods that can use \textit{in situ} data to optimize system parameters in an online fashion, selecting new parameters based on previous measurements, and subject to process constraints.

In this study, we focus on some key parameters that jointly affect the surface roughness of a print. The print speed and the amount of extruded material are the two main parameters that influence surface roughness. The \emph{print speed} $v_p$ is the speed of the extruder head during the deposition of a layer. The amount of extruded material is often characterized by the {extrusion multiplier} $e_m$, which multiplies the baseline extrusion amount (i.e. the theoretically needed amount of material according to extrusion modeling~\citep{aksoy2020control}) to produce a fine-tuned command for the extruder.

\subsection{Bayesian Optimization}

BO is an iterative strategy for the optimization of black-box and expensive-to-evaluate functions, often under performance constraints. In a general optimization problem, the objective function $f(\mathbf{x})$ and constraint function $c(\mathbf{x})$ are modeled by Gaussian process (GP) regression trained with data. The GP models can produce predictions of the functions (with the corresponding uncertainties) away from the training data points. In BO, the GP model predictions and uncertainty are used to select the input $\mathbf{x}^*$ at which to conduct the next evaluation. The evaluation results $f(\mathbf{x}^*)$ and $c(\mathbf{x}^*)$ are added to the available data set and the next optimization iteration takes place. The function returning the most valuable $\mathbf{x}$ to test depending on the already available data and models is called an \emph{acquisition function}. Numerous acquisition functions are presented in the BO literature~\citep{Hernandez-Lobato2016ASearch,Garrido-Merchan2019PredictiveConstraints,Gardner2014BayesianConstraints}. BO has been successfully used in numerous applications, such as manufacturing~\citep{maier2020self,guidetti2021plasma} or control under safety constraints~\citep{khosravi2022safety}. In this work, we use the BO algorithm studied in \citep{GuidettiAdvanced}, that was specifically designed for the configuration of advanced manufacturing processes such as FFF.

\subsection{Material}

The feedstock material we use in this work is LCP, which has been presented in \citep{gantenbein2018three} and is currently used for high-end applications by NematX AG\footnote{\texttt{https://nematx.ch}}. LCPs are composed of aromatic thermotropic polyesters. When heated above their melting temperature, these polyesters self-assemble into nematic domains (i.e.\ the molecules have their long axes arranged in parallel). In this spontaneous configuration, however, each coherent domain is oriented in a different and random direction, and no global molecular arrangement in the material is present. Extruding the material through a heated nozzle -- the typical deposition method in FFF -- has been shown to produce global alignment: the deformations and forces created by the extrusion process reorient the nematic domains in the extrusion direction. Upon exit from the nozzle, the aligned nematic domains are frozen in place by the rapid cooling caused by exposure to ambient temperature. After printing, the monomers are thus aligned in the axial direction of the deposited filament. 


\section{Problem Statement} \label{sec:problem}

The monomer alignment achieved in LCP FFF produces extraordinary mechanical properties, comparable to traditional but more complex fiber-reinforced materials. However, LCP is very sensitive to the parameters of the FFF deposition process. Studying similar polymers, it has been shown that, during the deposition of adjacent material lines, contact between an existing line and the nozzle printing the next line causes drag and subsequent misalignment in the previously deposited monomers~\citep{siqueira2017}. Clearly, reducing the fraction of aligned monomers lowers the mechanical performance of manufactured components. This effect has been shown experimentally to exist in LCP printing. For example, in the case of over-extrusion, the excess of deposited material is unable to achieve proper monomer alignment and disturbs the alignment of existing lines in a similar fashion to nozzle contact. Conversely, in the event of under-extrusion, the monomer alignment is not impacted, but the amount of deposited material is lower than what would be necessary to solidly fill the part, making the mechanical properties sub-optimal.

Thus, the print quality -- which can be quantified by layer inspection to detect over- or under-extrusion -- affects the performance of printed components. In this work, we propose to optimize the FFF of LCPs while using surface roughness as an easy-to-measure \textit{in situ} proxy for mechanical performance. The contributions of this work are
\begin{enumerate}
    \item the validation of an \textit{in situ} method for surface quality evaluation using a laser distance sensor,
    \item a study on the correlation between \textit{in situ} measured print quality and mechanical properties assessed via destructive testing, and
    \item the successful application of a sample-efficient optimization algorithm to the FFF of LCPs.
\end{enumerate}

\section{Methods} \label{sec:methods}

\subsection{Surface Quality Evaluation}

To evaluate the quality of the material deposition process, we propose to analyze the surface of each deposited layer while printing. We have modified a printer head to accommodate a compact laser triangulation sensor. The sensor returns the distance between the printer head and the point directly below it. When moving the printer head horizontally (i.e.\ in a plane parallel to the print bed), the sensor readings can be used to reconstruct the entire profile of the scanned surface (see e.g.~\cite{balta2021layer}). In Alg. \ref{alg:scan} we detail the steps required to produce an \textit{in situ} layer-by-layer scan of a part made of $N$ layers.

\begin{algorithm2e}[htbp]
\caption{\textit{In Situ} Layers Surface Scanning}\label{alg:scan}
\For{$k \gets 1$ to $N$}{
    Deposit layer $k$; \\
    Lift the printer head; \\
    Begin recording laser sensor measurements; \\
    Move the printer head horizontally and \emph{perpendicularly} to the print lines, to pass over one complete section of layer $k$ at constant traveling speed; \\
    End recording and save data from layer $k$;
}
\end{algorithm2e}

For each layer, we obtain a sequence of distance measurements associated with the position of the printer head. This data can be processed to draw an elevation profile of the layer section (cf.\ Sec.\ \ref{sec:res_surf}) or to compute a quantitative evaluation of the layer surface roughness.

We use the ISO 4287 profile parameter Ra to quantify the roughness of a profile. This is a commonly used texture parameter \citep{TOWNSEND} and is calculated as
\begin{equation} \label{eq:ra}
    \mathrm{Ra} = \frac{1}{K}\sum_{i = 1}^{K} |z_i - z_\mathrm{mean}| \,,
\end{equation}
where $K$ is the total number of distance measurements recorded in a layer, $z_i$ indicates each indexed measurement, and $z_\mathrm{mean}$ is the mean of all measurements. After computing the roughness Ra of each layer, we compute the average roughness across all layers to produce a part \emph{global roughness} $R$. In the rest of this work, we will use this quantity to assess the quality of printed samples.

\subsection{FFF Parameters Optimization}

In light of the connection between print quality and mechanical properties outline in Sec. \ref{sec:problem}, we propose to optimize the printing process by solving the following constrained optimization problem
\begin{equation} \label{eq:optprob}
    \begin{array}{rl}    
    \maxOp_{\{v_p,e_m\} \in \mathcal{X}}\!
    & \quad
    v_p
    \\\text{s.t.}
    & \quad
    R(v_p,e_m) \leq \lambda\,,
\end{array}
\end{equation}
where the print speed $v_p$ and the extrusion multiplier $e_m$ belong to a predefined set $\mathcal{X}$, and the global roughness of the printed samples is bounded by $\lambda$.

To solve Problem \eqref{eq:optprob}, we use the constrained BO algorithm introduced in \citep{GuidettiAdvanced}. The method was specifically developed for problems having a deterministic objective function (i.e. that can be directly computed from the decision variables $v_p$ and $e_m$) and an unknown constraint that can be evaluated only point-wise, by manufacturing a sample and analyzing it. As the manufacturing and analysis of a sample need to be completed before being able to select print parameters for the next sample, the process is time-consuming, which makes the use of sample-efficient optimization methods particularly appropriate.

The selected optimization algorithm is parameterized by the confidence threshold $\pi \in [0,1]$, which makes it switch between two different acquisition functions during the iterative optimization procedure. In short, when the confidence to find print parameters producing samples that respect the roughness constraint is larger than $\pi$, the algorithm will select parameters that increase the print speed drastically. Conversely, if the uncertainty is large and the algorithm confidence in meeting the constraints is smaller than $\pi$, it will propose print parameters that favor constraint satisfaction and improve the print speed only marginally. The choice of $\pi$ impacts the behavior of the optimization procedure. Small values of $\pi$ make the optimization aggressive, with large potential improvements in the print speed but few samples meeting the constraints. Setting $\pi$ to large values makes the procedure cautious, maximizing the fraction of printed samples that meet the constraints and optimizing the print speed slowly. All details about the method can be found in \citep{GuidettiAdvanced}.

\section{Results} \label{sec:res}

The results discussed in this section were obtained by manufacturing parts on a custom-built FFF machine actuated by Linax linear motors having an accuracy of \SI{1}{\micro\meter}, a maximum speed of \SI{2}{m/s} and a maximum acceleration of \SI{40}{m/s^2}. The parts were printed using LCP material fed to an E3D Hemera extruder with a nozzle diameter of \SI{0.4}{\milli\meter}.

\begin{figure}
\begin{center}
\hspace{0.2cm}
 \begin{minipage}[c]{0.15\columnwidth}
    \includegraphics[height=2.2in]{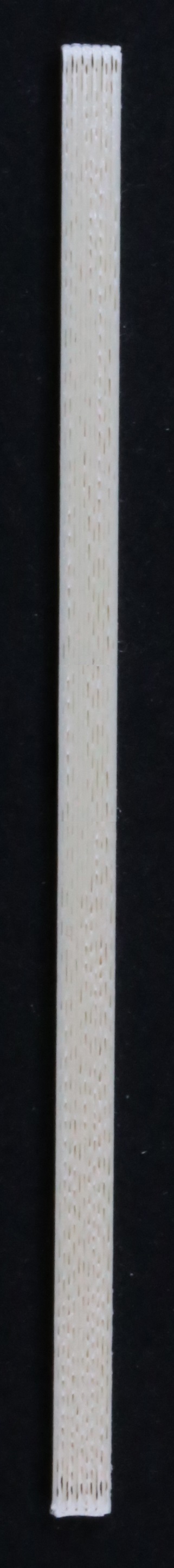}
  \end{minipage}
  \begin{minipage}[c]{0.8\columnwidth}
    \includegraphics{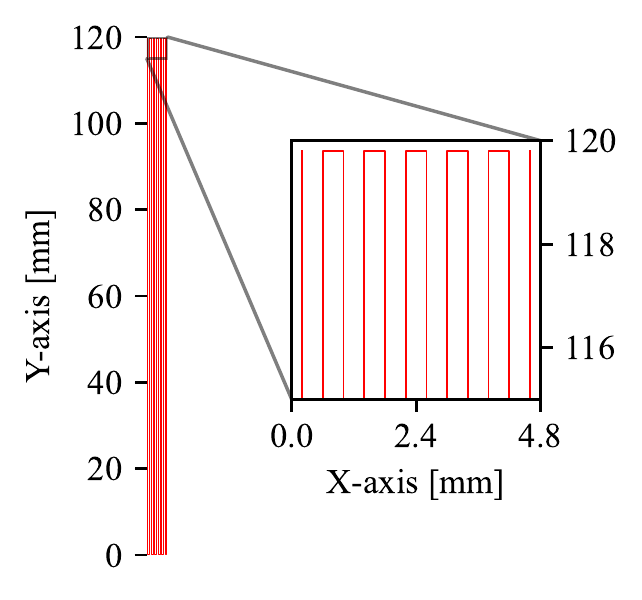} 
  \end{minipage}
\caption{Picture of a reference sample and illustration of the print trajectories used to manufacture it.} 
\label{fig:sample}
\end{center}
\end{figure}

During the optimization procedure, we used as a reference geometry a cuboid with length \SI{120}{mm}, width \SI{4.8}{mm} and height \SI{2}{mm} (cf. Fig. \ref{fig:sample}). It was printed by depositing parallel lines oriented in the direction of the maximum length of the cuboid. The line distance was set at \SI{0.4}{mm} (identical to the nozzle diameter) to produce a 100\% dense infill. The layer height was \SI{0.05}{mm}. The printer bed was kept at a temperature of \SI{180}{\celsius} while the nozzle temperature was set to \SI{320}{\celsius}.

\subsection{Surface Quality Evaluation and Validation} \label{sec:res_surf}

We mounted a Micro-Epsilon ILC1220-10 laser distance sensor on the printer head, paired with an RS422 interface. The sensor has a repeatability of \SI{1}{\micro\meter} and a measurement range of \SI{1}{cm}. We fixed the measuring rate to \SI{1}{\kilo\hertz} and selected a scan speed (i.e.\ the translation speed of the printer head during surface scanning) of \SI{500}{mm/\minute}.

\begin{figure}
\begin{center}
\includegraphics{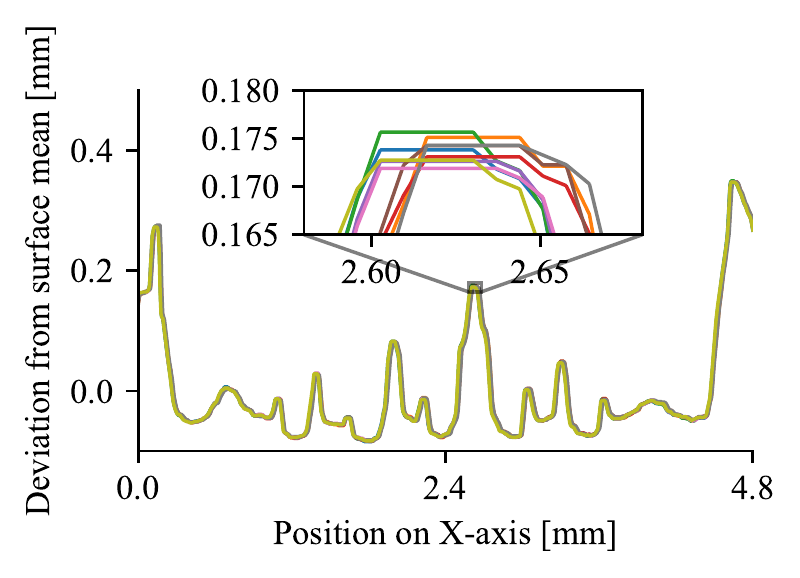}    
\caption{Surface profiles produced by scanning a reference sample nine times at the same location. The inset shows the repeatability of the process: what appears as a single curve is in fact nine curves effectively superimposed on each other.} 
\label{fig:repeat}
\end{center}
\end{figure}

We first analyzed the repeatability of the surface measurement by printing one high-roughness reference sample and then scanning its top surface repeatedly. As explained in Alg. \ref{alg:scan}, a measurement is conducted by moving the laser perpendicularly to the print lines (i.e.\ across the width of the sample), from one edge to the other. This motion was repeated nine times. To verify that the measurements are independent of the scanning direction, we moved the head back and forth, with five scans taking place from left to right and four scans from right to left. We overlapped all measurements to produce the surface profiles shown in Fig. \ref{fig:repeat}.
For each pass of the sensor over the reference sample, we computed the corresponding roughness Ra. We then quantified the properties of this set of evaluations in Table \ref{tab:repeat}.
\begin{table}[htbp]
\centering
\caption{Statistical properties of the nine roughness evaluations obtained by repeatedly scanning a reference sample at the same location.} \label{tab:repeat}
\renewcommand{\arraystretch}{1.3}
\begin{tabular}[h]{@{}l r@{}}
\toprule
Minimum roughness & \SI{70.88}{\micro\meter} \\
Maximum roughness & \SI{71.19}{\micro\meter} \\
Mean roughness $\mu$ & \SI{70.99}{\micro\meter} \\
Roughness standard deviation $\sigma$ & \SI{0.09}{\micro\meter} \\
Coefficient of variation $\sigma/\mu$ & \SI{0.13}{\percent} \\
\bottomrule
\end{tabular}
\label{tab:slice_alignment}
\end{table}
By looking at Fig. \ref{fig:repeat}, and particularly at the detail of one peak of the profile, we observe that the distance measurements lie in a \SI{4}{\micro\meter} range. This is to be expected, as the repetitions are affected by the accuracy of the machine actuators and of the laser sensor. The observed variations are significantly smaller than the profile signal, and consequently do not impact significantly the quality of the roughness computation. Table \ref{tab:repeat} indicates that the changes in roughness are smaller than the sensor repeatability (\SI{1}{\micro\meter}), suggesting that the measurement procedure is effective. The coefficient of variation of \SI{0.13}{\percent} confirms that the roughness evaluation is extremely repeatable and the results produced with one laser pass can be trusted with high confidence.

\begin{figure}
\begin{center}
\includegraphics[trim={0 1.2cm 0 1.2cm},clip]{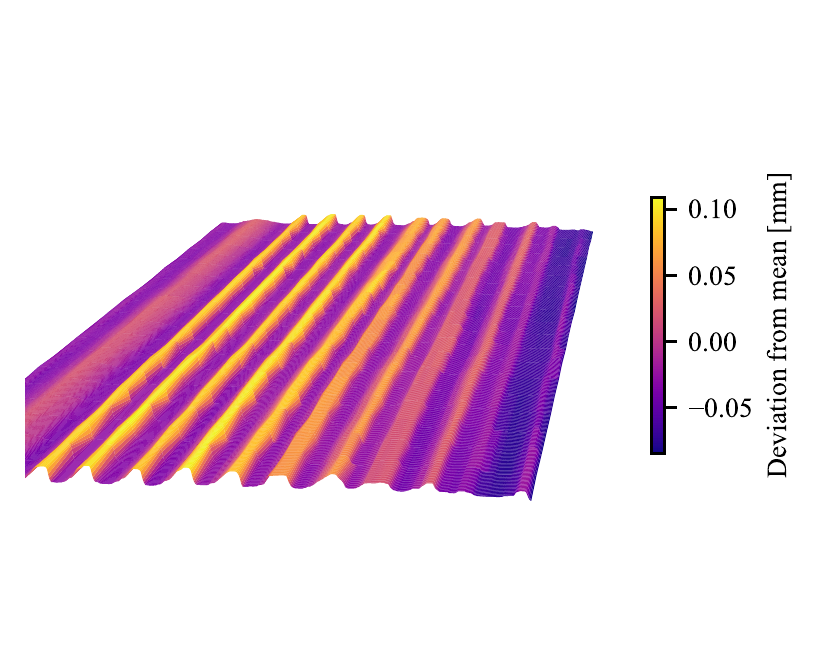}    
\caption{3D surface profile of the center of a sample, reconstructed from a raster laser scan.} 
\label{fig:scan_3D}
\end{center}
\end{figure}

To confirm that the measurements obtained from one section of the sample can be considered meaningful for the entire surface, we conducted 20 edge-to-edge scans at different locations. Each pass was translated from the previous one (in the sample length direction) by \SI{1}{mm} to produce a raster scan perpendicular to the print direction. We then processed the data by joining the points belonging to neighboring passes to produce the 3D surface profile of Fig. \ref{fig:scan_3D}.
The quality and continuity of the reconstruction confirm that little to no variation happens as we observe different sections of the sample. This supports the usage of a single laser pass over the printed sample to evaluate the roughness and consequentially the print quality of the entire layer.

\begin{figure}
\begin{center}
\includegraphics{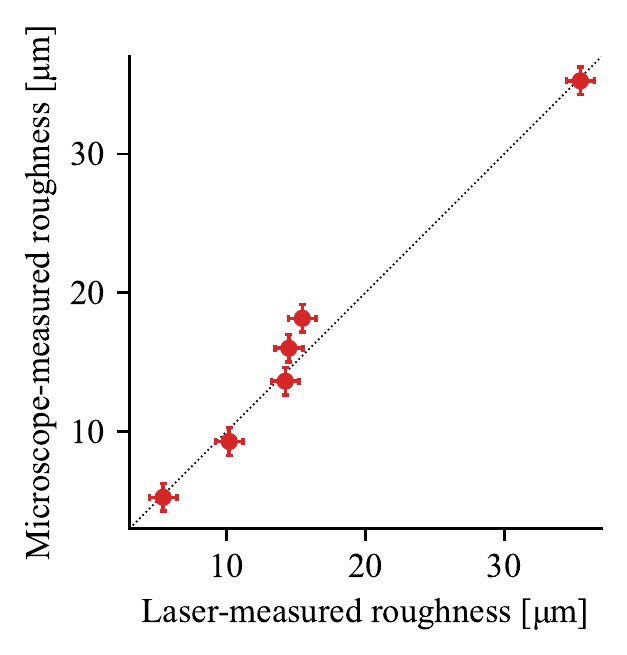}    
\caption{Comparison of the roughness evaluations of six different samples, conducted with the proposed \textit{in situ} laser scan and with a digital microscope surface scan. The error bars indicate the instrument's accuracy. The identity line is shown in dotted style.} 
\label{fig:micr_comp}
\end{center}
\end{figure}

Finally, we validated the measurements obtained with our \textit{in situ} sensors with those produced offline using a Keyence VHX 5000 digital microscope set at a $200\times$ magnification. The microscope software can compute automatically the roughness of a scanned surface according to \eqref{eq:ra}. We printed six reference samples with different print parameters to produce variations in the top surface roughness. We then scanned them with the laser sensor (while printing them) and with the microscope (after removing them from the printer and moving them to the microscope). We compared the Ra computed with our method and the one obtained from the microscope software (i.e.\ the validation ground truth) and plotted the results in Fig. \ref{fig:micr_comp}.
The results allow us to validate the laser sensor as a viable and accurate online alternative to more time-consuming procedures such as microscope evaluation. The roughness obtained with a single laser pass over the sample follows the ground truth with very little deviation and can be confidently used to evaluate the surface quality \textit{in situ}.

\subsection{Mechanical Properties Correlation} \label{sec:res_corr}

In this section, we describe the experimental evaluation of the correlation between the global roughness measure and the mechanical performance.
The samples were evaluated following the \emph{Standard Test Method for Tensile Properties of Polymer Matrix Composite Materials} by \citep{astm_prot}. The parts were clamped (at a pressure of \SI{600}{\kilo\pascal}) in a Zwick Z020 testing machine and underwent a tensile test conducted in the direction of the print lines. The clamps were moved apart at a speed of \SI{2}{mm/\min} and the procedure was interrupted when the part broke. By measuring the applied force and corresponding elongation continuously, we could compute Young's modulus $E$ and the ultimate tensile strength (UTS) of each sample.

We printed seven samples with different roughness $R$. To achieve this, we fixed the print speed to $v_p = \SI{350}{mm/s}$ and set the extrusion multiplier to seven different levels $e_m = \{0.7, 0.8, 0.9, 1.0, 1.1, 1.3, 1.5\}$, one per each sample. We scanned each printed layer by a single pass at the end of the deposition (in accordance with the findings of Sec.\ \ref{sec:res_surf}) to compute the roughness $R$ of each sample and then tested its mechanical properties.

\begin{figure}
\begin{center}
\includegraphics{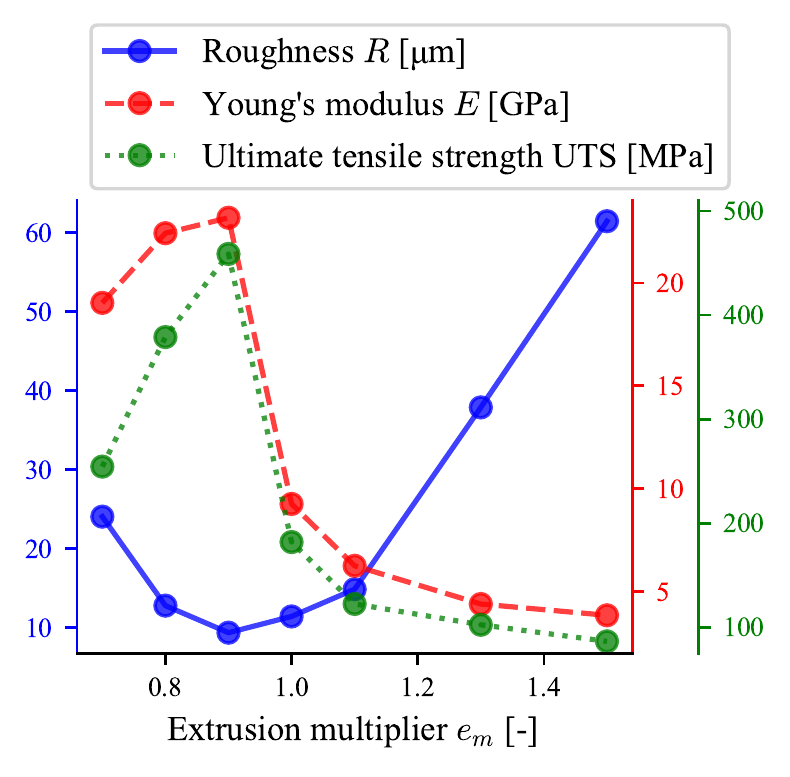}    
\caption{Roughness, Young's modulus, and ultimate tensile strength of seven samples printed with different extrusion multipliers.} 
\label{fig:corr}
\end{center}
\end{figure}
The collected data is shown in Fig.\ \ref{fig:corr}.
It is immediately evident that the lowest roughness correspond the best mechanical properties. The findings confirm the results from the literature and support the claim that poor print quality causes a reduction in mechanical performance. In the under-extrusion domain ($e_m = \{0.7, 0.8\}$), the sample infill is not complete: the voids between deposited lines increase the measured roughness, and the reduction in material density weakens the part. The drop in performance is however much more pronounced in the event of over-extrusion ($e_m = \{1.0, 1.1, 1.3, 1.5\}$). The excessive deposition of material interferes with the monomers' alignment, making them rearrange in random directions and thus negatively affecting the mechanical properties. The ridges produced on the surface by over-extrusion are easily detectable and significantly increase the roughness. In general, these results confirm that the \textit{in situ} roughness measurements are an easy-to-obtain and reliable alternative to time-consuming destructive testing and that they can be used for process optimization. 

\subsection{FFF Parameters Optimization}

We used the algorithm by \citep{GuidettiAdvanced} to solve Problem \eqref{eq:optprob}. The algorithm proceeds iteratively by recommending a pair of extrusion multiplier and print speed to use during the manufacturing of a sample. The user prints the sample while measuring the roughness online. The global roughness data is then returned to the algorithm, which uses the measurements to compute a new pair of print parameters for the next part.

We set the constraint $\lambda = \SI{10}{\micro\meter}$, which (based on the results shown in Fig.\ \ref{fig:corr}) corresponds to prints with extremely low roughness and thus peak mechanical properties. We call the prints respecting this roughness constraint \emph{feasible}. To respect the equipment limitations, we bounded the extrusion multiplier to $0.5 < e_m < 1.5$ and the print speed to $\SI{10}{mm/s} < v_p < \SI{500}{mm/s}$; these constraints encode the set $\mathcal{X}$ in \eqref{eq:optprob}. Finally, we initialized the models used by the optimization algorithm with the small data set collected during the study of Sec.\ \ref{sec:res_corr}. The study comprises 31 iterations (i.e.\ printed samples), of which the first 17 were conducted with $\pi = 0.4$ (balanced approach) and the remaining 14 with $\pi = 0.1$ (aggressive approach).

\begin{figure}
\begin{center}
\includegraphics{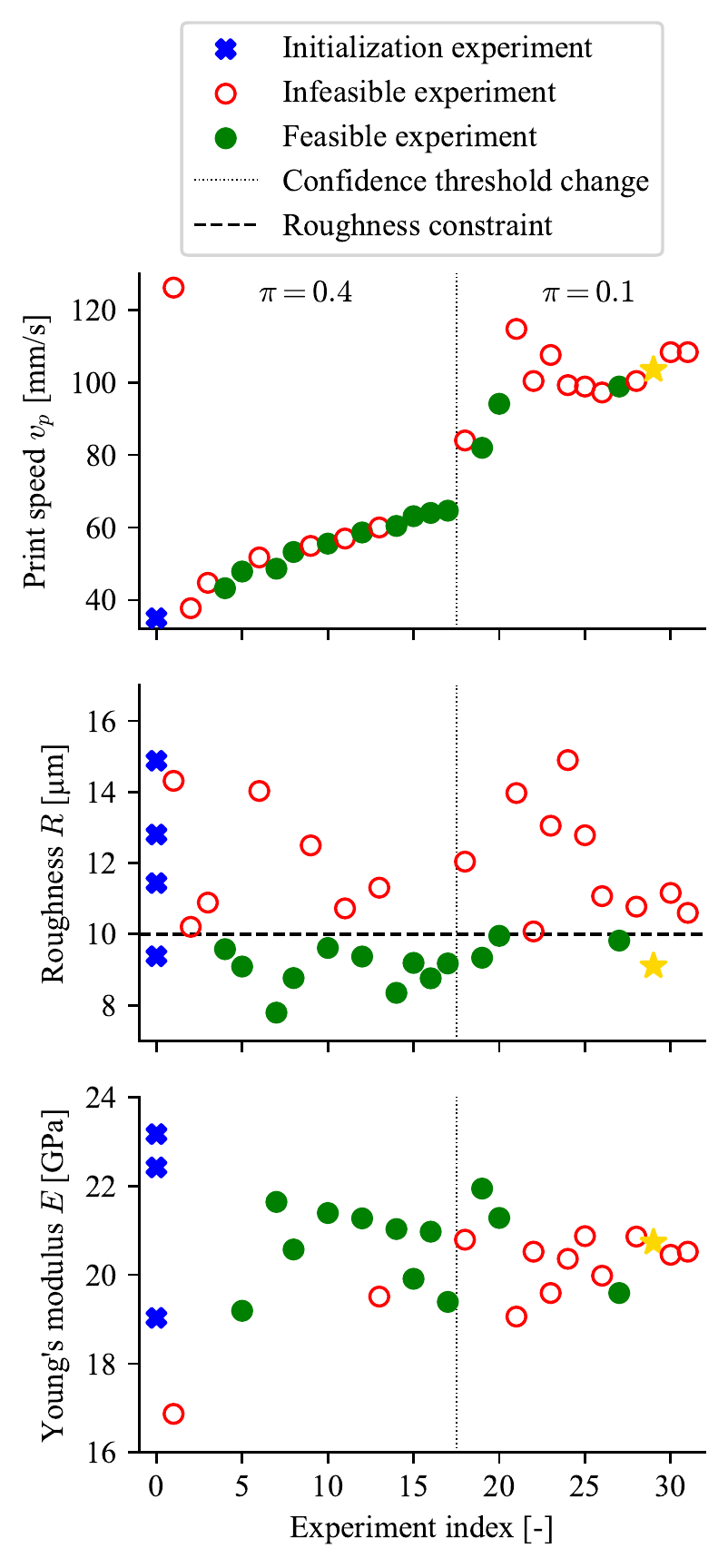}    
\caption{Experiments for the optimization of FFF, conducted with two different values of $\pi$. The golden star denotes the best feasible point found in the experiments. The Young's modulus was evaluated for 25 samples only due to testing time limitations.} 
\label{fig:opt}
\end{center}
\end{figure}

Figure \ref{fig:opt} shows the trace of the BO procedure. 
The different results produced by setting $\pi = 0.4$ and $\pi = 0.1$ are clearly visible. In the initial phase, where the approach was more cautious, \SI{59}{\percent} of the samples respect the roughness constraint. To this corresponds a slow but constant improvement in print speed, from the initial $v_p = \SI{35}{mm/s}$ to $v_p = \SI{64.4}{mm/s}$. This behavior is ideally suited to the manufacturing of small batches of usable prototypes (as most prints meet the requirements), during which the print process parameters are being continuously optimized (as the print speed keeps increasing). In the final phase of the experiment, where the approach was more aggressive, the algorithm attempted to sharply increase the print speed. This came at the cost of feasibility: only \SI{29}{\percent} of the samples respect the roughness constraint. However, by the time the optimization was interrupted, it was possible to find a set of parameters producing feasible prints at a speed of $v_p = \SI{103.5}{mm/s}$. This represents a $3\times$ improvement on the initialization print speed that was previously standard for LCP manufacturing set by subject matter experts.

The mechanical properties evaluation was based on Young's modulus only: a relevant fraction of the samples was so strong that the fracture would take place at the clamping site, rendering the UTS data unusable. However, the results in \citep{gantenbein2018three} confirm that Young's modulus and UTS are directly correlated, making the evaluation of Young's modulus sufficient to draw conclusions on the mechanical performance of a sample. The results achieved during optimization confirm the claim that low surface roughness corresponds to better properties. Clearly, as the optimization targets the low roughness region and the worst samples have $R = \SI{15}{\micro\meter}$, we do not expect to observe large drops or variations in Young's modulus, like those seen in Fig. \ref{fig:corr}. However, considering the feasible samples only we find an average Young's modulus $E = \SI{20.52}{\giga\pascal}$, while the non-feasible prints have a lower $E = \SI{19.95}{\giga\pascal}$. The best sample discovered in the optimization can be manufactured $3\times$ faster than the initialization samples, while still having a high-performing Young's modulus $E = \SI{20.73}{\giga\pascal}$.

\section{Conclusion}

We propose a data-driven method for the automated selection and optimization of process parameters in FFF. We test and validate an \textit{in situ} laser distance sensor for the evaluation of surface roughness. We then correlate the surface quality of manufactured parts with their mechanical properties obtained via destructive testing. Finally, we use the surface roughness measurement to guide the optimization of print speed and extrusion rate used in LCP manufacturing. We were able to find a set of print parameters that maintained the mechanical properties of a sample part while reducing the time required for manufacturing by \SI{66}{\percent} when compared to the state of praxis.  

\begin{ack}
We gratefully acknowledge the support of NematX AG in all the experiments conducted for this research.
\end{ack}

\bibliography{ifacconf}  

\begin{thebibliography}{4}
\providecommand{\natexlab}[1]{#1}
\providecommand{\url}[1]{\texttt{#1}}
\providecommand{\urlprefix}{URL }
\expandafter\ifx\csname urlstyle\endcsname\relax
  \providecommand{\doi}[1]{doi:\discretionary{}{}{}#1}\else
  \providecommand{\doi}{doi:\discretionary{}{}{}\begingroup
  \urlstyle{rm}\Url}\fi

\bibitem[{Able(1956)}]{Abl:56}
Able, B. (1956).
\newblock Nucleic acid content of microscope.
\newblock \emph{Nature}, 135, 7--9.

\bibitem[{Able et~al.(1954)Able, Tagg, and Rush}]{AbTaRu:54}
Able, B., Tagg, R., and Rush, M. (1954).
\newblock Enzyme-catalyzed cellular transanimations.
\newblock In A.~Round (ed.), \emph{Advances in Enzymology}, volume~2, 125--247.
  Academic Press, New York, 3rd edition.

\bibitem[{Keohane(1958)}]{Keo:58}
Keohane, R. (1958).
\newblock \emph{Power and Interdependence: World Politics in Transitions}.
\newblock Little, Brown \& Co., Boston.

\bibitem[{Powers(1985)}]{Pow:85}
Powers, T. (1985).
\newblock Is there a way out?
\newblock \emph{Harpers}, 35--47.

\end{thebibliography}


\begin{thebibliography}{19}
\providecommand{\natexlab}[1]{#1}
\providecommand{\url}[1]{\texttt{#1}}
\providecommand{\urlprefix}{URL }
\expandafter\ifx\csname urlstyle\endcsname\relax
  \providecommand{\doi}[1]{doi:\discretionary{}{}{}#1}\else
  \providecommand{\doi}{doi:\discretionary{}{}{}\begingroup
  \urlstyle{rm}\Url}\fi

\bibitem[{Aksoy et~al.(2020)Aksoy, Balta, Tilbury, and
  Barton}]{aksoy2020control}
Aksoy, D., Balta, E.C., Tilbury, D.M., and Barton, K. (2020).
\newblock A control-oriented model for bead cross-sectional geometry in fused
  deposition modeling.
\newblock In \emph{2020 American Control Conference (ACC)}, 474--480. IEEE.

\bibitem[{{ASTM International}(2014)}]{astm_prot}
{ASTM International} (2014).
\newblock \emph{Standard Test Method for Tensile Properties of Polymer Matrix
  Composite Materials}.
\newblock \doi{10.1520/D3039\_D3039M-08}.

\bibitem[{Balta et~al.(2021)Balta, Tilbury, and Barton}]{balta2021layer}
Balta, E.C., Tilbury, D.M., and Barton, K. (2021).
\newblock Layer-to-layer stability of linear layerwise spatially varying
  systems: Applications in fused deposition modeling.
\newblock \emph{IEEE Transactions on Control Systems Technology}, 29(6),
  2517--2532.

\bibitem[{Dey and Yodo(2019)}]{dey2019systematic}
Dey, A. and Yodo, N. (2019).
\newblock A systematic survey of fdm process parameter optimization and their
  influence on part characteristics.
\newblock \emph{Journal of Manufacturing and Materials Processing}, 3(3), 64.

\bibitem[{Gantenbein et~al.(2018)Gantenbein, Masania, Woigk, Sesseg, Tervoort,
  and Studart}]{gantenbein2018three}
Gantenbein, S., Masania, K., Woigk, W., Sesseg, J.P., Tervoort, T.A., and
  Studart, A.R. (2018).
\newblock Three-dimensional printing of hierarchical liquid-crystal-polymer
  structures.
\newblock \emph{Nature}, 561(7722), 226--230.

\bibitem[{Gardner et~al.(2014)Gardner, Kusner, Xu, Weinberger, and
  Cunningham}]{Gardner2014BayesianConstraints}
Gardner, J.R., Kusner, M.J., Xu, Z., Weinberger, K.Q., and Cunningham, J.P.
  (2014).
\newblock {Bayesian optimization with inequality constraints}.
\newblock \emph{31st International Conference on Machine Learning, ICML 2014},
  3, 2581--2591.

\bibitem[{Garrido-Merch{\'{a}}n and
  Hern{\'{a}}ndez-Lobato(2019)}]{Garrido-Merchan2019PredictiveConstraints}
Garrido-Merch{\'{a}}n, E.C. and Hern{\'{a}}ndez-Lobato, D. (2019).
\newblock {Predictive Entropy Search for Multi-objective Bayesian Optimization
  with Constraints}.
\newblock \emph{Neurocomputing}, 361, 50--68.
\newblock \doi{10.1016/j.neucom.2019.06.025}.

\bibitem[{Guidetti et~al.(2021)Guidetti, Rupenyan, Fassl, Nabavi, and
  Lygeros}]{guidetti2021plasma}
Guidetti, X., Rupenyan, A., Fassl, L., Nabavi, M., and Lygeros, J. (2021).
\newblock Plasma spray process parameters configuration using sample-efficient
  batch bayesian optimization.
\newblock In \emph{2021 IEEE 17th International Conference on Automation
  Science and Engineering (CASE)}, 31--38. IEEE.

\bibitem[{Guidetti et~al.(2022)Guidetti, Rupenyan, Fassl, Nabavi, and
  Lygeros}]{GuidettiAdvanced}
Guidetti, X., Rupenyan, A., Fassl, L., Nabavi, M., and Lygeros, J. (2022).
\newblock Advanced manufacturing configuration by sample-efficient batch
  bayesian optimization.
\newblock \emph{IEEE Robotics and Automation Letters}, 7(4), 11886--11893.
\newblock \doi{10.1109/LRA.2022.3208370}.

\bibitem[{Hern{\'{a}}ndez-Lobato et~al.(2016)Hern{\'{a}}ndez-Lobato, Gelbart,
  Adams, Hoffman, and Ghahramani}]{Hernandez-Lobato2016ASearch}
Hern{\'{a}}ndez-Lobato, J.M., Gelbart, M.A., Adams, R.P., Hoffman, M.W., and
  Ghahramani, Z. (2016).
\newblock {A general framework for constrained {B}ayesian optimization
  optimization using information-based search}.
\newblock \emph{Journal of Machine Learning Research}, 17, 1--53.

\bibitem[{Khosravi et~al.(2022)Khosravi, Koenig, Maier, Smith, Lygeros, and
  Rupenyan}]{khosravi2022safety}
Khosravi, M., Koenig, C., Maier, M., Smith, R.S., Lygeros, J., and Rupenyan, A.
  (2022).
\newblock Safety-aware cascade controller tuning using constrained bayesian
  optimization.
\newblock \emph{IEEE Transactions on Industrial Electronics}.

\bibitem[{Maier et~al.(2020)Maier, Rupenyan, Bobst, and
  Wegener}]{maier2020self}
Maier, M., Rupenyan, A., Bobst, C., and Wegener, K. (2020).
\newblock Self-optimizing grinding machines using gaussian process models and
  constrained bayesian optimization.
\newblock \emph{The International Journal of Advanced Manufacturing
  Technology}, 108(1), 539--552.

\bibitem[{Mohamed et~al.(2016)Mohamed, Masood, and
  Bhowmik}]{mohamed2016mathematical}
Mohamed, O.A., Masood, S.H., and Bhowmik, J.L. (2016).
\newblock Mathematical modeling and fdm process parameters optimization using
  response surface methodology based on q-optimal design.
\newblock \emph{Applied Mathematical Modelling}, 40(23-24), 10052--10073.

\bibitem[{Peng et~al.(2014)Peng, Xiao, and Yue}]{peng2014process}
Peng, A., Xiao, X., and Yue, R. (2014).
\newblock Process parameter optimization for fused deposition modeling using
  response surface methodology combined with fuzzy inference system.
\newblock \emph{The International Journal of Advanced Manufacturing
  Technology}, 73(1), 87--100.

\bibitem[{Siqueira et~al.(2017)Siqueira, Kokkinis, Libanori, Hausmann, Gladman,
  Neels, Tingaut, Zimmermann, Lewis, and Studart}]{siqueira2017}
Siqueira, G., Kokkinis, D., Libanori, R., Hausmann, M.K., Gladman, A.S., Neels,
  A., Tingaut, P., Zimmermann, T., Lewis, J.A., and Studart, A.R. (2017).
\newblock Cellulose nanocrystal inks for 3d printing of textured cellular
  architectures.
\newblock \emph{Advanced Functional Materials}, 27(12), 1604619.
\newblock \doi{10.1002/adfm.201604619}.

\bibitem[{Sood et~al.(2009)Sood, Ohdar, and Mahapatra}]{sood2009improving}
Sood, A.K., Ohdar, R., and Mahapatra, S.S. (2009).
\newblock Improving dimensional accuracy of fused deposition modelling
  processed part using grey taguchi method.
\newblock \emph{Materials \& design}, 30(10), 4243--4252.

\bibitem[{Townsend et~al.(2016)Townsend, Senin, Blunt, Leach, and
  Taylor}]{TOWNSEND}
Townsend, A., Senin, N., Blunt, L., Leach, R., and Taylor, J. (2016).
\newblock Surface texture metrology for metal additive manufacturing: a review.
\newblock \emph{Precision Engineering}, 46, 34--47.
\newblock \doi{https://doi.org/10.1016/j.precisioneng.2016.06.001}.

\bibitem[{Turner and Gold(2015)}]{turner2015review}
Turner, B.N. and Gold, S.A. (2015).
\newblock A review of melt extrusion additive manufacturing processes: Ii.
  materials, dimensional accuracy, and surface roughness.
\newblock \emph{Rapid Prototyping Journal}.

\bibitem[{Wankhede et~al.(2020)Wankhede, Jagetiya, Joshi, and
  Chaudhari}]{wankhede2020experimental}
Wankhede, V., Jagetiya, D., Joshi, A., and Chaudhari, R. (2020).
\newblock Experimental investigation of fdm process parameters using taguchi
  analysis.
\newblock \emph{Materials Today: Proceedings}, 27, 2117--2120.

\end{thebibliography}

\end{document}